\documentclass[a4paper,prb,superscriptaddress,citeautoscript,twocolumn,twoside,showpacs]{revtex4}

\usepackage{amsmath}
\usepackage{graphicx}
\usepackage{amssymb}
\usepackage{amstext}
\usepackage{color}

\usepackage[T1]{fontenc}
\usepackage[utf8]{inputenc}
\usepackage{float}

\newcommand{\ket}[1]{|\!#1\rangle}
\newcommand{\eg}{{{e.g.}}}
\newcommand{\ie}{{{i.e.}}}

\newcommand{\Hl}{\leftrightarrow}
\newcommand{\V}{\updownarrow}
\newcommand{\D}{\mathbin{\rotatebox[origin=c]{45}{$\updownarrow$}}}
\newcommand{\A}{\mathbin{\rotatebox[origin=c]{135}{$\updownarrow$}}}
\newcommand{\Lc}{\circlearrowleft}
\newcommand{\R}{\circlearrowright}
\newcommand{\citeref}[1]{~{[\!\!\citenum{#1}]}}

\begin{document}

\title{Experimental quantum forgery of quantum optical money}

\author{Karol Bartkiewicz}\email{bark@amu.edu.pl}
\affiliation{Faculty of Physics, Adam Mickiewicz University,
PL-61-614 Pozna\'n, Poland}
\affiliation{RCPTM, Joint Laboratory
of Optics of Palack\'y University and Institute of Physics of
Academy of Sciences of the Czech Republic, 17.~listopadu 12, 772
07 Olomouc, Czech Republic}
\affiliation{CEMS, RIKEN, 351-0198 Wako-shi, Japan}

\author{Anton\'{i}n \v{C}ernoch}
\affiliation{Institute of Physics of Czech Academy of Sciences,
Joint Laboratory of Optics of PU and IP AS CR, 17.~listopadu 50A,
772 07 Olomouc, Czech Republic}

\author{Grzegorz Chimczak}
\affiliation{Faculty of Physics, Adam Mickiewicz University,
PL-61-614 Pozna\'n, Poland}

\author{Karel Lemr}
\affiliation{RCPTM, Joint Laboratory
of Optics of Palack\'y University and Institute of Physics of
Academy of Sciences of the Czech Republic, 17.~listopadu 12, 772
07 Olomouc, Czech Republic}

\author{Adam Miranowicz}
\affiliation{Faculty of Physics, Adam Mickiewicz University,
PL-61-614 Pozna\'n, Poland}
\affiliation{CEMS, RIKEN, 351-0198 Wako-shi, Japan}

\author{Franco Nori}
\affiliation{CEMS, RIKEN, 351-0198 Wako-shi, Japan}
\affiliation{Department of Physics, The University of Michigan,
Ann Arbor, MI 48109-1040, USA}

\begin{abstract}
Unknown quantum information cannot be perfectly copied (cloned).
This statement is the bedrock of quantum technologies and quantum
cryptography, including the seminal scheme of Wiesner's quantum
money\cite{Wiesner83}, which was the first quantum-cryptographic
proposal. Surprisingly, to our knowledge, quantum money has not
been tested experimentally yet. Here, we experimentally revisit
the Wiesner idea, assuming a banknote to be an image encoded in the
polarization states of single photons. We demonstrate that it is
possible to use quantum states to prepare  a banknote that cannot
be ideally copied  without making the owner aware of only
unauthorized actions. We provide  the security conditions for 
quantum money by investigating the physically-achievable
limits on the fidelity of 1-to-2 copying of arbitrary sequences of
qubits. These results can be applied as a security measure in
quantum digital right management.
\end{abstract}


\date{\today}
\maketitle

%============================================================

%\section{Introduction}

The seminal proposal of quantum money by Wiesner\cite{Wiesner83}
(see also Ref.~[\!\!\citenum{Bennett82}]), followed by the
introduction of quantum key distribution (QKD) protocols by Bennet
and Brassard\cite{BB84} and by Ekert\cite{Ekert91}, have triggered
a breathtaking interest and progress not only in quantum
cryptography but, in general, in quantum information over the last
three decades. It is not surprising that
Refs.~[\!\!\citenum{BB84,Ekert91}] on QKD are among the most often
cited works in quantum information and both quantum and classical
cryptography. Moreover, various commercial implementations of QKD
protocols (for a recent review see~[\!\!\citenum{Lo14}]), together
with quantum random-number generators and the  D-Wave
machine (see, e.g.,~[\!\!\citenum{Jones13}])
% OR: D-Wave computer (which however causes much controversy about its quantumness\cite{Jones13}),
are probably the only commercial applications of quantum
information and quantum optics up to now\cite{Georgescu12}.
Although, various protocols of quantum money have already been
proposed (see, e.g.
Refs.~[\!\!\citenum{Burham01,Barnum02,Tokunaga03,Mosca10,Farhi10,Lutomirski10b,Pastawski12,Aaronson11,Molina13}]),
this interest cannot be compared with the immense popularity and
applicability of QKD
(see~Refs.~[\!\!\citenum{Karol13prl,Sasaki14,Takesue15}] as an
example of recent and fundamental achievements). This is partially
because there have not been, to our knowledge, any experimental
realizations of quantum money performed yet. Here, we report not
only an experimental implementation of quantum money but also an
experimental attempt to its forgery using optimal cloning
machines.

Our experimental work basically describes one-by-one attacks on
each single qubit. In the quantum money scheme, however,
eavesdroppers, in principle, can access every qubit at once. So,
they can globally access multiple qubits and can seek superior
attacks using such global access. This could be a reason why there
has not been a known representative work for the experiment of
attacking quantum money, because this would need to treat numerous
qubits and difficult global controls of their quantum
states. The attacks presented in this work are less distinguished
from quantum cloning itself or the attack for BB84 quantum key
distribution. Thus, collective or coherent attacks on multiple
qubits simultaneously can, in principle, optimize the attacker's
strategy. This is, nevertheless, considerably more demanding if
not impossible with the current state of experimental quantum
information processing. In this paper, we investigate a more
accessible form of attack based on individual cloning which, in
our view, represents a realistic threat for near-future quantum
communications, including quantum money schemes.

Any information can be encoded as a sequence of zeros and ones.
This sequence can also be represented using a set of single
photons prepared in the horizontal and vertical polarization
states. The polarization states of a photon can be described as a
superposition of the two orthogonal polarization states, \ie,
\begin{equation}\label{eq:psi}
|\psi\rangle=\cos\frac{\theta}{2}|\Hl\,\rangle+\mathrm{e}^{i\phi}\sin\frac{\theta}{2}|\V\,\rangle,
\end{equation}
where the angles $\theta$ and $\phi$ are the spherical coordinates
of this qubit on the Bloch sphere, while $\Hl$ and $\V$ denote
horizontal and vertical polarisations, respectively. For each such
state there exists an orthogonal state
\begin{equation}\label{eq:ortpsi}
|\psi_\perp\rangle=\sin\frac{\theta}{2}|\Hl\,\rangle-\mathrm{e}^{i\phi}\cos\frac{\theta}{2}|\V\,\rangle.
\end{equation}
Any pair of such orthogonal states can be used to encode logical
values 0 and 1. Without knowing what particular states have been
used (\ie, without knowing $\theta$ and $\phi$), there is no way
of telling (with certainty) what logical value is associated with
the photon.

Any attempt of gaining this information from the photon will
disturb its polarization state and damage the information.
Therefore, using photons to transmit sensitive information appears
to be a promising idea.
% CITE: quantum cryptography papers, etc.
In the simplest scenario, the sequence of polarized photons is
associated with a set of numbers indicating the correct
measurement bases. These latter sequence needs to be confidential.
If this sequence would be intercepted together with the sequence
of photons, the quantum information could be read and reproduced
at will. First, by deterministically distinguishing between
$|\psi\rangle$ and $|\psi_\perp\rangle$ associated with the bit
values $0$ and $1$, respectively. Next, by reproducing the
detected state.

Therefore, the advantages provided by this kind of quantum
communication are limited to protocols, where a trusted arbiter
checks the validity of a given sequence of qubits. Thus, the
sequence of qubits can be used, \eg,  as  one-time passwords
(tokens)\cite{Pastawski12}
%CITE: quantum tokens
or arbitrated quantum currency\cite{Wiesner83}.
%CITE: Wisesner.
However, some research has been conducted in order to eliminate
the need for an arbiter in the quantum currency schemes
\cite{Farhi10, Aaronson11}.

Currently, tokens are widely applied as an extra layer of
security, \eg, in a two-step authentication protocols used in
social media services or Internet banking etc. While the classical
tokens are sensitive to being copied, the quantum tokens cannot be
delivered to two or more users at the same time without disturbing
a given quantum dataset
\cite{Wiesner83,Aaronson11,Pastawski12,Molina13}.
%CITE: nocloning, tokens

It is claimed today that the security of our data is as good as
its passwords. In the following text we discuss how to generate
and check the security of the best tokens allowed by the laws of
nature. The quantum passwords cannot be copied nor viewed without
damaging them. However, quantum data are prone to noise and some
level of noise has to be tolerated in order to harness the
benefits of quantum technologies.

The quantum tokens can also be used as quantum money. The idea of
quantum money goes back to Wiesner
%CITE: Wiesner
\cite{Wiesner83} who proposed to embed a sequence of qubits into
banknotes that would be verified by banks. This was the first idea
of quantum cryptography introduced already in the early 1970s and
eventually published in 1983\cite{Wiesner83,Gisin02RMP}.

In order to be able to verify the money, a bank would attach
information about the banknote serial number as classical
information. This pioneering  idea evolved over the last decades
to more practical protocols, which are shown to be more secure and
less demanding on the participating parties of a quantum currency
system \cite{Farhi10,Aaronson11}. However, all the protocols face
the problem of decoherence that makes the quantum banknotes to be
usable for a limited amount of time, even if the currency is
represented as a sequence of photons \cite{Pastawski12}, which can
have exceptionally-long coherence times.

Photons are robust to decoherence, because they do not usually
interact with each other. Moreover, if the string of photons is
handled properly it can last in a coherent state long enough to be
useful in some financial transactions. Let us consider a
transaction, where quantum money is withdrawn at the speed of
light from a bank by an authorized user as a sequence of photons
that arrives at a payment terminal, which allows its user to
redirect the money to any other payment terminal. The final user
sends the sequence to the bank together with an account number,
where the money is to be  stored. Lossless transmission of photons
is impossible. Therefore, banks would have to accept large enough
parts of incomplete quantum banknotes and issue new ones. The same
is done nowadays if a banknote is damaged or a small part of it is
missing. The communication between the payment terminals cannot be
wiretapped without damaging this quantum money. Thus, this quantum
money scheme (QMS) allows for some anonymity if the addresses of
the terminals are not assigned to a specific person and there is
at least one terminal used between the initial and final users.
However, the money could be signed without damaging it using, \eg,
the approach discussed in \mbox{Ref.~\citeref{Karol14}}.

\begin{figure*}
\includegraphics[width=14.5cm]{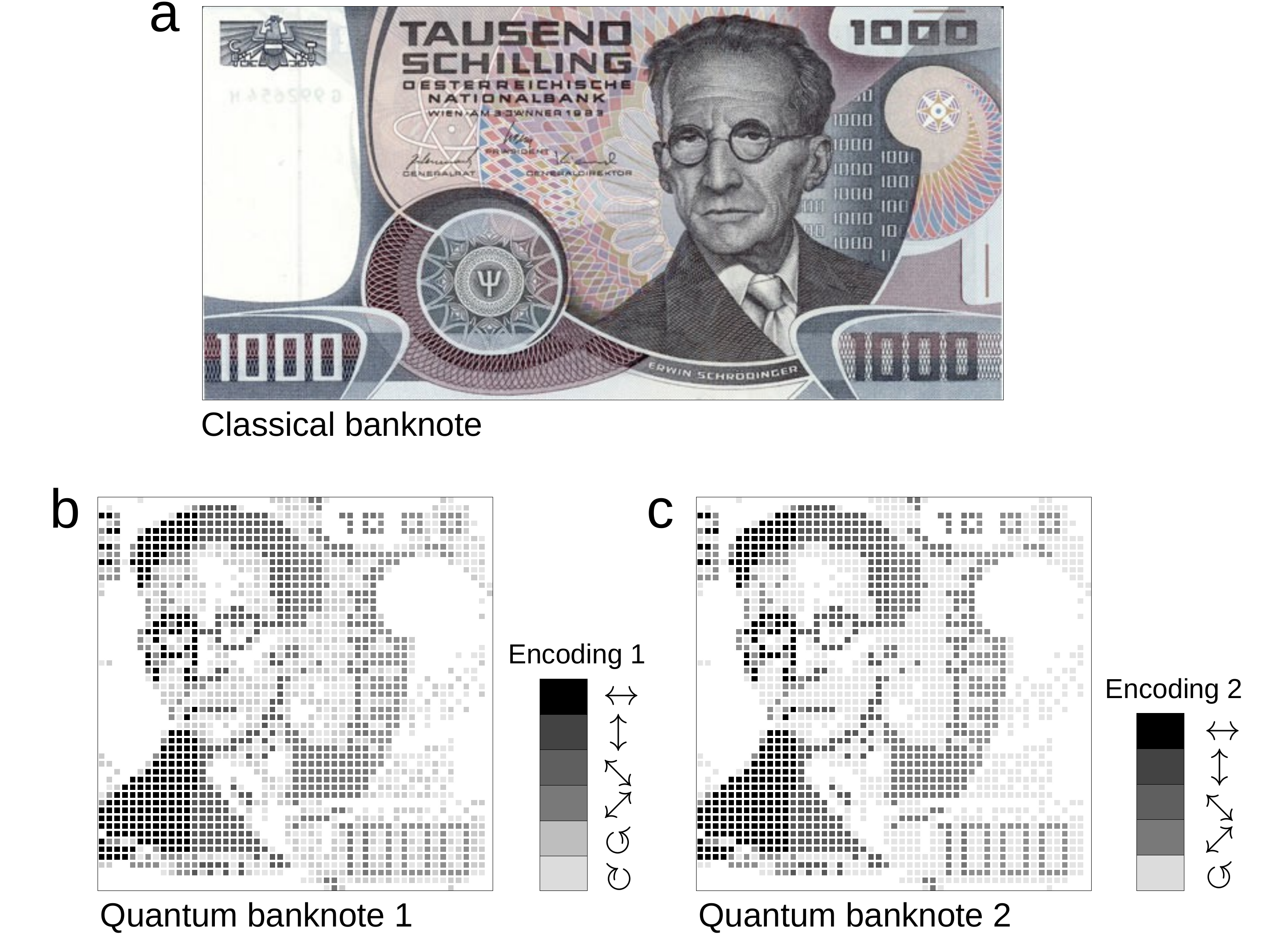}
\caption{ \label{fig:token}  An illustrative example of (a) a
classical banknote. In panels (b)
and (c) the simplified banknote from panel (a) 
with the decreased number of colors and resolution
is encoded
experimentally in two ways to form two examples of quantum
banknotes. The symbols used here correspond to different linear
($\Hl{}$, $\V{}$, ...) and circular ($\Lc{}$ and $\R{}$) photon
polarisations (as explained in the main text). Note that the white
regions in (b) and (c) correspond to the lack of photons.}
\end{figure*}

Perfect copying of quantum information is impossible
%CITE: nocloning theorem
\cite{Wooters82,Dieks82,Aaronson11},
%CITE: Complexity-Theoretic No-Cloning Theorem
but as it was shown in various works,  we can copy partially-known
quantum information with very high fidelity. If we are going to
clone some qubits more often than others, we can use a generic
distribution function $g(\theta,\phi)$ to describe this intent.
The higher the value of $g$, the more frequent cloning of the
specific qubit is. This distribution function satisfies the
following normalization condition
\begin{equation}
\int_{\Omega}g(\theta,\phi)\,\mathrm{d}\Omega=1,
\end{equation}
where
$\mathrm{d}\Omega\equiv\sin\theta\,\mathrm{d}\phi\,\mathrm{d}\theta$
and $\Omega$ is the full solid angle. The distribution $g$ can be
arbitrary, but until now only highly-symmetric distributions have
been analysed (see, \eg,~\mbox{\citeref{Fiurasek03,Karol09,Karol10}}
and references therein). Therefore, one can be under the impression
that this optimal cloning problem can be solved only for a
highly-symmetric class of distributions. However, as we show
below, we are in principle able to always find an optimal cloning
machine corresponding to any randomly generated quantum tokens or
banknotes. Note that the most secure tokens are the ones with the
highest entropy. The same applies here, because the lowest average
cloning fidelity, corresponding to the case most resistant to
cloning attacks, is achieved for a uniform distribution $g$, which
has the highest possible entropy. However, while generating
quantum money of a finite size at random, it is hard to ensure
each time the perfect entropy. Therefore, in practice, we could
deal with any qubit distribution function $g$ that could be
potentially known to the counterfeiter. In particular, there exist
qubit distributions $g$ made of a weighted sum of two
Dirac's delta functions at any antipodes of the Bloch sphere. In
this special case, the problem is reduced to the classical case of
standard digital tokens. This is because these particular
functions tell us that there are only two states sent that could
be discriminated deterministically. Quantum money of this kind
should  obviously be avoided.

Let us briefly review the main possible attack scenarios. Without
any knowledge about the token, the counterfeiter can use the
universal quantum cloner \cite{UC96}.
%CITE: universal cloner
If the states, appearing in the qubit sequence, are known but
their order is unknown, the attacker can apply a specialized
optimal quantum cloning machine. This is equivalent to the
situation in which the attacker has some information about the
money statistics, but does not know the sequence of qubits itself.
The results of such an attack can be seen in Fig.~\ref{fig:token}.
Unfortunately, if the attacker knows the sequence of bases,
the quantum money (tokens) can be perfectly copied. \\

%=======================================================================================

%\section{Noise tolerance vs. security}

\begin{figure}
\includegraphics[width=8cm]{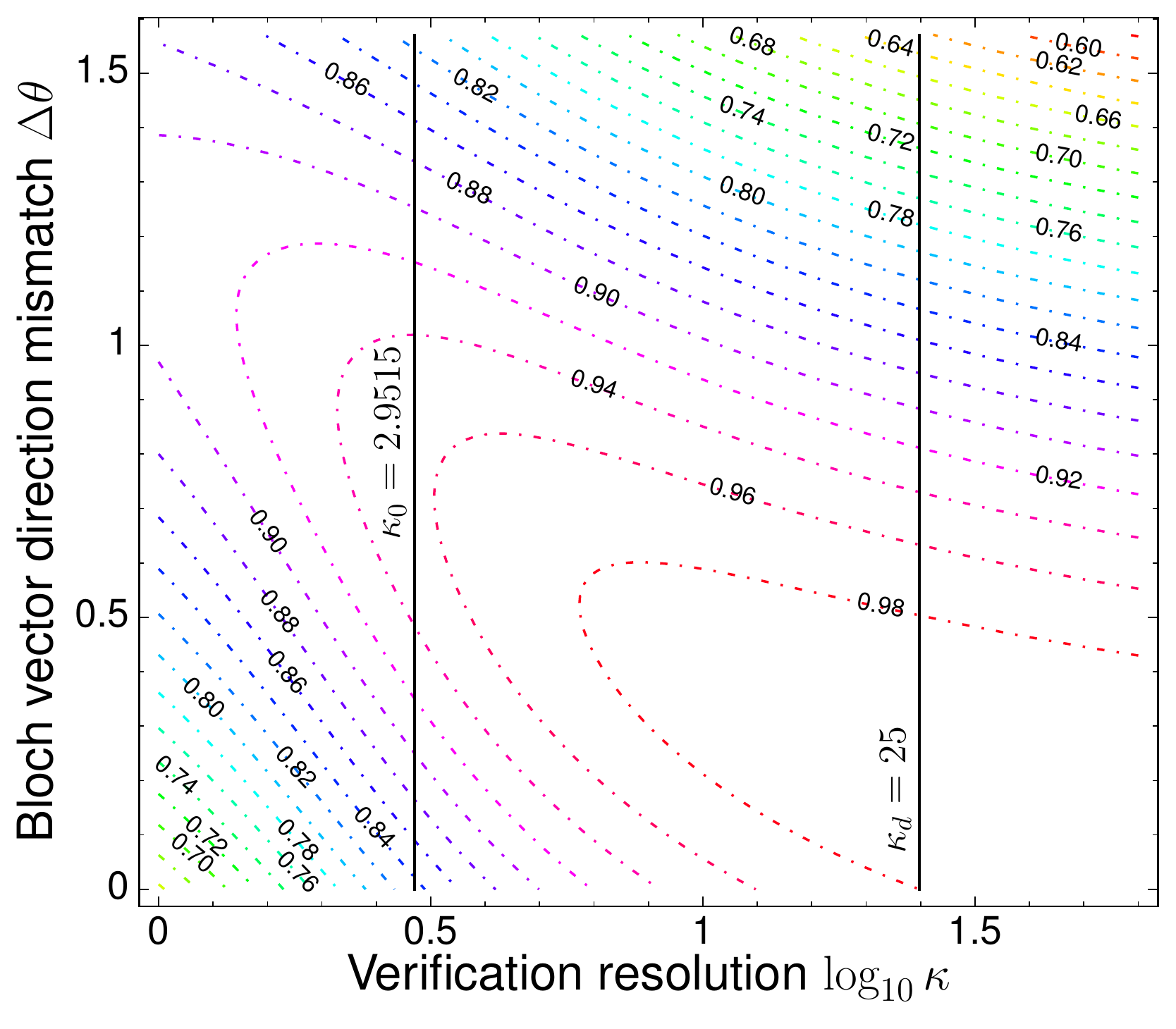}
\caption{\label{fig:kappa0} Contour plot showing how the
probability of detecting a qubit of an unknown state
$|\psi(\theta,\phi)\rangle$ for money verification depends on the
imperfect choice of the measurement direction $\Delta\theta$ and
the photon verification (or discrimination) resolution $\kappa$.
Specifically, this probability is equivalent to the fidelity
$F_{\mathrm{proc}}(\Delta\theta,\kappa)$ for which the Bloch
vector is rotated with respect to its correct orientation by
$\Delta\theta$ for a given value of $\kappa$. Note that the
probability does not depend on $\theta$ or $\phi$, but only on
$\Delta\theta$, which measures the angle between the original and
rotated Bloch vectors. The solid black lines mark two specific
values of $\kappa$: $\kappa_0=2.9515$ describes the minimal
resolution needed to detect an attack with an optimal universal
cloning machine and $\kappa_d=25$ corresponds to the resolution
reached in our experiment. Note that the shape of the depicted
relation depends on the dispersion function of the detector. Here,
this function is chosen as the von Mises--Fisher distribution.}
\end{figure}

\noindent {\bf Noise tolerance versus security.} Let us estimate the
level of noise tolerance needed for a quantum token to be
validated in realistic conditions and compare it to the level of
noise introduced by a given optimal quantum cloning. By doing
so, we will limit the class of distributions associated with
acceptable tokens. We assume that a counterfeiter can replace the
noisy communication channel with a less noisy one and perform a
quantum man-in-the-middle attack with an optimal quantum cloning
machine. An equivalent assumption is that the counterfeiter is a
party in the QMS. Finding the optimal cloning transformation for a
given $g$ is a semi-definite programming problem\cite{F1,Aud02}.
Such problem can be described as a task of finding a semi-definite
operator $\hat{\chi}$  (a cloning map) describing the copying
process that maximizes the average single-copy fidelity $F$. Such
operator is isomorphic to a completely positive trace-preserving
(CPTP) map \cite{Jam72}. The average single-copy fidelity for an
arbitrary distribution (for symmetric $1\to2$ cloning) can be
expressed \cite{Karol09,Karol10} as
\begin{equation}
F=\frac{1}{2}\int_{\Omega}g(\theta,\phi)\left(F_{0}+F_{1}\right)\mathrm{d}\Omega,
\end{equation}
where the fidelities of copying a particular qubit for the first
and second clones are
\begin{equation}
F_{0}=\mathrm{Tr}\left[(\hat{\rho}^{\mathrm{T}}\otimes\hat{\rho}\otimes\hat{\openone})\hat{\chi}\right]\,\textrm{and}\,
F_{1}=\mathrm{Tr}\left[(\hat{\rho}^{\mathrm{T}}\otimes\hat{\openone}\otimes\hat{\rho})\hat{\chi}\right]\,,
\end{equation}
where $\hat{\rho}=|\psi\rangle\langle\psi|$, $\mathrm{T}$ stands
for transposition, and $\hat{\openone}$ is the single-qubit
identity operator. The density matrices of both clones are
identical and they read
$\rho_{i}=\mathrm{Tr}_{\mathrm{in},i\oplus1}\left[(\hat{\rho}^{\mathrm{T}}\otimes{\openone}^{\otimes2})\hat{\chi}\right]$,
where we calculate the partial trace over the input qubit and one
of the two clones ($\oplus$ stands for sum modulo $2$).

The average single-copy fidelity written in a compact form reads
\begin{equation}\label{eq:Fopt}
F=\mathrm{Tr}\left(\hat{R}\hat{\chi}\right).
\end{equation}
In order to find the optimal cloning map $\hat{\chi}$, one needs
to compute the $\hat{R}$ operator defined as
\begin{equation}
\hat{R}=\frac{1}{2}\int_{\Omega}g(\theta,\phi)\hat{\rho}^{\mathrm{T}}\otimes\left(\hat{\openone}\otimes\hat{\rho}+\hat{\rho}\otimes\hat{\openone}\right)\mathrm{d}\Omega.
\end{equation}
Remarkably,  we show in the Methods that this operator depends
only on its  five expansion coefficients of $g$ in the basis of
spherical harmonics, regardless of the exact form of $g$. The
optimal map $\hat{\chi}$ is found by maximizing $F$ in
Eq.~(\ref{eq:Fopt}) for a given $\hat{R}$ with the optimization
algorithm described in \citeref{F1} (see also
Refs.~\citeref{Fiurasek03,Karol09,Karol10,Karol13prl,Karol14amp}).

The output distribution $g_{\mathrm{out}}$ of the cloned qubits
will differ from $g$, because perfect cloning is impossible. Each
cloning machine prepares a perfect clone  (\ref{eq:psi}), with
probability equal to the fidelity $F_{i}$, and an
orthogonal state (\ref{eq:ortpsi}), with  probability
$1-F_{i}$. Thus, the distribution $g_{\mathrm{out}}(\theta,\phi)$
of the cloned qubit states can be expressed as
\begin{eqnarray}
g_{\mathrm{out}}(\theta,\phi)&=&F_{i}(\theta,\phi)g(\theta,\phi)\\
&&+\left[1-F_{i}(\theta+\pi,\phi+\pi)\right]g(\theta+\pi,\phi+\pi).\nonumber
\end{eqnarray}
There is no difference between $g$ and $g_{\mathrm{out}}$, if the
function is symmetric with respect to inverting the directions of
the Bloch sphere. This includes the scenarios both for the best
case (a uniform qubit distribution) and the worst case (a sequence
of distinguishable states). The class of such distributions
defines the so-called mirror phase-covariant cloner (or
cloning) (MPCC)\cite{Karol09}. Note that the MPCC is a
generalization of the phase-covariant cloners (PCCs), which enable
optimal copying of a qubit state from the equator of the Bloch
sphere\cite{Bruss00} or other states on the Bloch sphere with a
definite angle $\theta$\cite{Karimipour02,Fiurasek03} (see the
Supplementary Material\cite{Supplement} for more details about the
MPCC and PCC). The output distribution cannot be used directly to
quantify the quality of the clones, because it does not carry 
the information about the order of states in a given sequence.

The analysed sequence would usually contain some additional noise
due to small random polarization rotations caused by various
imperfections. These include state preparation,
distribution, storage, and finally delivery and analysis. In
practice, all these imperfections lead to the average sequence
fidelity $F_\mathrm{pass}<1$ with respect to the ideally-performed
qubit preparation, storage, and detection steps.

For simplicity, we assume that all the enlisted protocol elements
are perfect, except the final step of our state analysis. If this
final step is the polarization analysis of single photons with
standard detectors and a polarization beam splitter, we have
$F_\mathrm{pass}\approx 98\%$. Here, we model the joint dispersion
of the transmission channel and the state verification with
respect to the target polarization by the spherical dispersion
model on a sphere given by the von~Mises--Fisher  distribution
\cite{Fisher53}
%CITE: von Mises -- Fisher
(\ie, the Gaussian distribution on a sphere)
\begin{equation}
f(\kappa,\alpha) = \frac{\exp(\kappa\cos\alpha)}{2\pi
I_0(\kappa)},
\end{equation}
which is the probability density function of any qubit prepared in
a target state given by its Bloch vector being rotated by an angle
$\alpha$. The level of concentration of the density function
around the state vector $|\psi\rangle$  is given by the parameter
$\kappa$. The density function is normalized with the modified
Bessel function $I_0(\kappa)$\cite{Arfken85}. From this model it
follows that the probability of detecting a qubit described by the
density matrix $\rho=|\psi\rangle \langle \psi|$ is equivalent to
the average fidelity (\ref{eq:Fopt}) and is given by
\begin{equation}
F_{\mathrm{proc}}(\rho_{},\kappa)=\int_{0}^{\pi}\int_{0}^{2\pi}
f(\kappa,\alpha)\langle \mu |\rho_{}|\mu\rangle\,
\mathrm{d}\delta\, \mathrm{d}\alpha \,,
\end{equation}
where $|\mu\rangle=\cos(\frac{\theta-\alpha}{2})|\psi\rangle +
\exp(i\delta)\sin(\frac{\theta-\alpha}{2})|\psi_\perp\rangle$. For
example, our direct calculations for $\alpha=0$ lead to
$F_{\mathrm{proc}}(\theta,\kappa) = [2\kappa \cos\theta\cosh\kappa
+ \pi\kappa I_1(\kappa)\sin\theta +
2(\kappa-\cos\theta)\sinh\kappa)]/(4\kappa\sinh\kappa),$ where
$I_1$ is the modified Bessel function\cite{Arfken85}. Thus, for
the QMS to be feasible, we need to accept those sequences with 
fidelity
$F_\mathrm{pass}=F_\mathrm{proc}(|\psi\rangle\langle\psi|,\kappa_{0})
$. Hence, $\kappa_0$ describes the minimum resolution required to
reveal an attack using a cloner with a given value of
$F_\mathrm{pass}$. The value of  $\kappa_{0}$ can be derived
numerically from the fixed value of $F_\mathrm{pass}$
corresponding to the fidelity of polarization analysis. For a
single qubit, we can use the following security condition $
F_{\mathrm{proc}}(\rho_{i},\kappa) <F_\mathrm{pass},$ where now
$\kappa$ describes the dispersion of the channel used by the
counterfeiter to deliver the copied sequence. If this condition is
satisfied, the counterfeiter cannot cheat the verification
process. The verification process is performed on the full
sequence of qubits. Therefore, any verification process that
allows for some implementation imperfections should depend on the
average verification fidelity. For a  long sequence of cloned
qubits this average fidelity is
\begin{eqnarray}
\bar{F}_{i}(\kappa)&=&\int_{\Omega}g(\theta,\phi)
F_{\mathrm{proc}}(\rho_{i},\kappa)\mathrm{d}\Omega\,,
\end{eqnarray}
whereas for the verification threshold reads as
\begin{equation}
\bar{F}_{\mathrm{pass}}(\kappa_0)=\int_{\Omega}
g(\theta,\phi){F}_{\mathrm{pass}}(\theta,\phi)\mathrm{d}\Omega.
\end{equation}
These values can be obtained by projecting the delivered quantum
banknote on the associated sequence of bases. These can be
approximated as the ratios of the number of the correctly
projected states to the number of the conclusive state
projections. A quantum banknote passes the verification process if
$\bar{F}_{i} > \bar{F}_{\mathrm{pass}}.$ These quantities
(used in this inequality) depend implicitly on the choice of $g$
as the quality of the optimally-counterfeited state depends on
$g$, specifically on its five expansion coefficients in terms of
spherical harmonics, \ie, five real numbers that could be
estimated by the counterfeiter after measuring some random parts of the
banknote.  Thus, in the following text, we assume that $g$ is
publicly known. We demonstrate experimentally that this weakness
could be exploited by a counterfeiter.

Let us consider the situation where the security threshold is
given by a theoretical value of $\bar{F}_{i}$, where
$\kappa\to\infty$, which does not take into account the threat of
the counterfeiter using the knowledge about $g$. In this case, one
would naively assume that the  forgery cannot lead to the fidelity
$\bar{F}_i$ exceeding $5/6$, corresponding to the fidelity of the
universal cloning machine \cite{UC96}. It would appear that using
the security threshold of $\bar{F}_{\mathrm{pass}}=5/6$ might be a
good idea, as it makes the QMS more  robust against errors. This
means that one could naively allow  the resolution of the
verification process  $\kappa_0$ to be as small as
$\kappa_0=2.9515$. This value is obtained from
$F_\mathrm{proc}(0,\kappa_0)=[\kappa_0 \cosh\kappa_0  +
(\kappa_0-1)\sinh\kappa_0]/(2\kappa_0\sinh\kappa_0)=5/6$. To
illustrate that this could be a problem, let us imagine that we
verify qubits described by the Bloch vectors rotated by an angle
$\Delta\theta$ from the Bloch vectors of the expected states. In
Fig.~\ref{fig:kappa0}, we see that the measured fidelity
$F_{\mathrm{proc}}(\Delta\theta,\kappa_0)$ would be seemingly
above the security threshold even for $\Delta\theta\approx\pi/2$,
which means that the verification process would recognize a large
volume of pure states as valid. However, it would not accept the
states for which the Bloch vectors are rotated by more that
$90^\circ$ from the target Bloch vectors. In this regime, we are
approaching the situation where any state prepared in a  basis,
which is unbiased with respect to the verification basis, would
pass the verification process. The counterfeiter can guess the
conjugate basis correctly with  probability $2/3$ and choose
the correct state in the matching basis with probability
$1/6$. This means that $83\%$ of an arbitrary banknote prepared by
the counterfeiter is accepted and the QMS is broken. Fortunately,
this is not exactly the case as
$F_{\mathrm{proc}}(\pi/2,2.9515)=0.8115<5/6$. Note that this could
be dangerous if the dispersion of the state verification would not
be described with the von Mises--Fisher distribution, but with
some similar function. Thus, for the low resolution regime of
$\kappa_0\approx 2.9515$ the full characterization of the
verification setup is required in order to exclude this classical
attack.

The detection resolution $\kappa_d$ of a given experimental setup
should be as large as possible. In our experiment we achieved
$\kappa_d=25$, which is obtained from
$F_\mathrm{proc}(0,\kappa_d)=[\kappa_0 \cosh\kappa_d  +
(\kappa_d-1)\sinh\kappa_d]/(2\kappa_d\sinh\kappa_d)=0.98$. Even if
the detection resolution is perfect $\kappa_0\to\infty$, the
quantum money can be counterfeited  using a specialized quantum
cloner optimized for $g$.
In the following section we illustrate this  with an experiment.\\

%\section{Experimental quantum forgery}

\begin{figure}
\includegraphics[width=8cm]{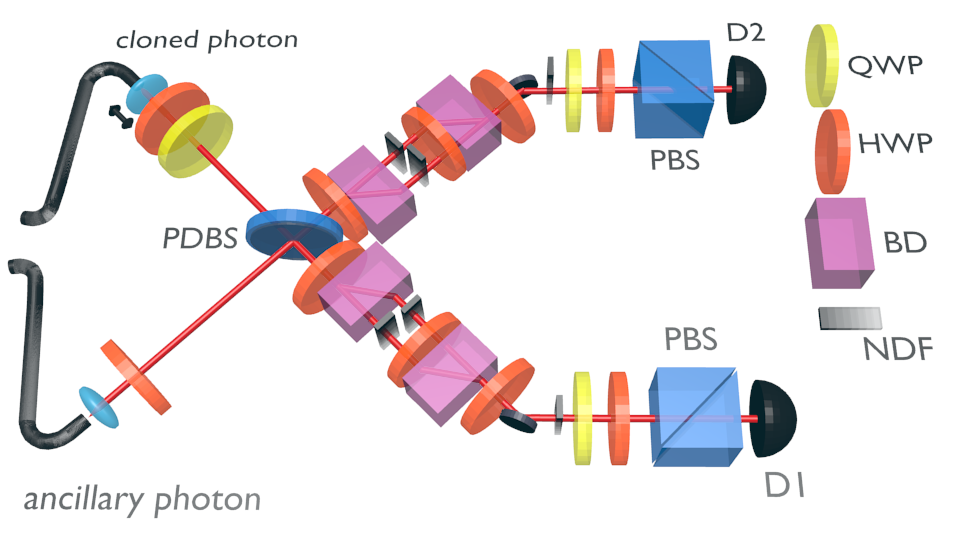}
\caption{\label{fig:setup} Experimental setup for cloning quantum
bankotes. Components are labelled as follows: HWP is half-wave
plate, QWP is quarter wave-plate, PDBS is polarisation
dependent-beam splitter, PBS is polarising beam splitter, BD is
beam divider, NDF is neutral density filter, and D is single
photon detector. A successful cloning and verification of a qubit
from a given sequence is registered as a simultaneous detection
event at the two detectors. }
\end{figure}

\begin{figure*}
\includegraphics[width=14.5cm]{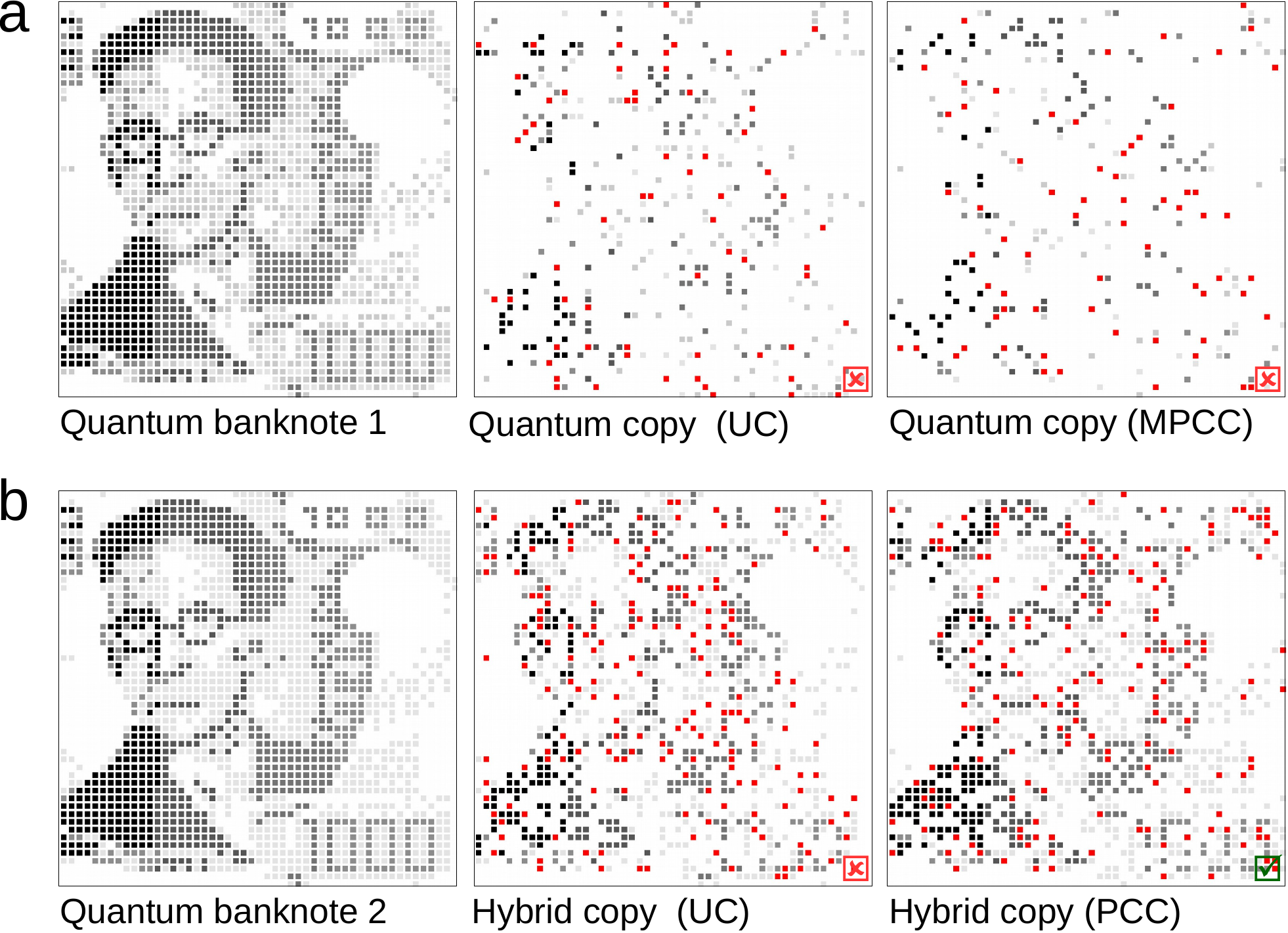}
\caption{ \label{fig:token2} Experimental quantum banknotes
1 (a) and 2 (b) are copied probabilistically with an optimal $1$-to-$2$ 
linear optical cloning machine shown in
Fig.~\ref{fig:setup} and subsequently verified. This device
can be tuned to implement, in special cases, the universal quantum
cloner (UC), the phase-covariant cloner (PCC), and the mirror
phase-covariant cloner (MPCC). Note that the white regions in
quantum banknotes, or their copies, correspond to either a lack of
photons or the cases where the cloning process failed to deliver
one photon per banknote. One observes that the copies, which are
provided with the best possible cloning machines, are noisy and,
thus, the sequences of qubits are damaged (shown in red). The
performance of a given cloning process depends on the statistics of
photon polarisations. Thus, the copies of quantum banknote~1 (a)
obtained by an optimal purely-quantum cloner (the UC and
MPCC) fail the verification. The copies of banknote~2 (b)
obtained by an optimal hybrid (i.e., quantum-classical) cloner
fail the verification if the UC is used, but pass the
verification if the PCC is applied.}
\end{figure*}

\begin{figure}
\includegraphics[width=8.5cm]{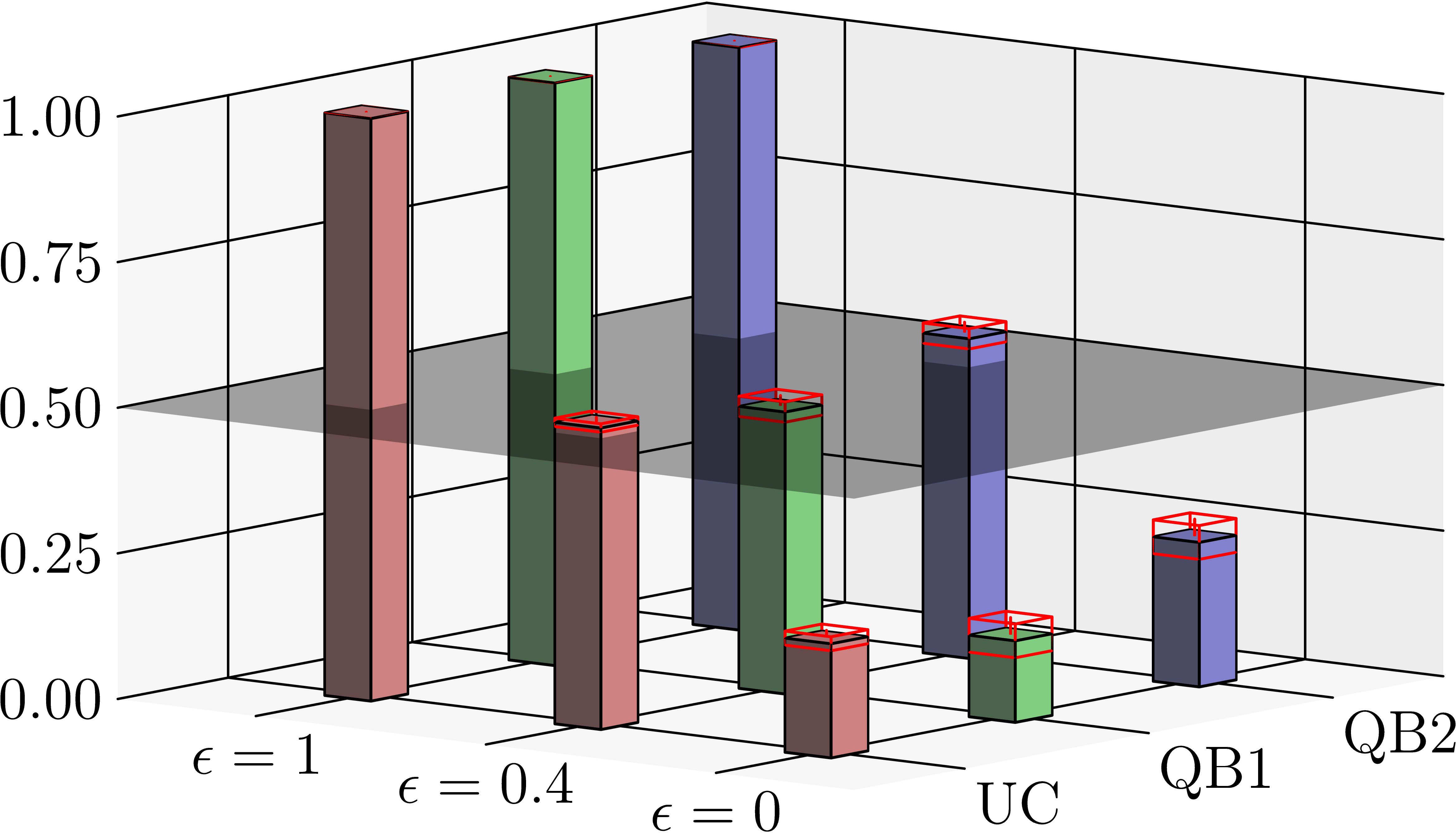}
\caption{\label{fig:probs} Success probabilities of the cloning processes for quantum banknotes~$1$ and $2$ (QB1 and QB2), and optimal universal cloning (UC). 
The red frames show the error bars of the measured probabilities. The grey surface shows the minimum cloning efficiency needed to output on average more cloned photons than the input photons.}
\end{figure}

\begin{figure}
\includegraphics[width=8.5cm]{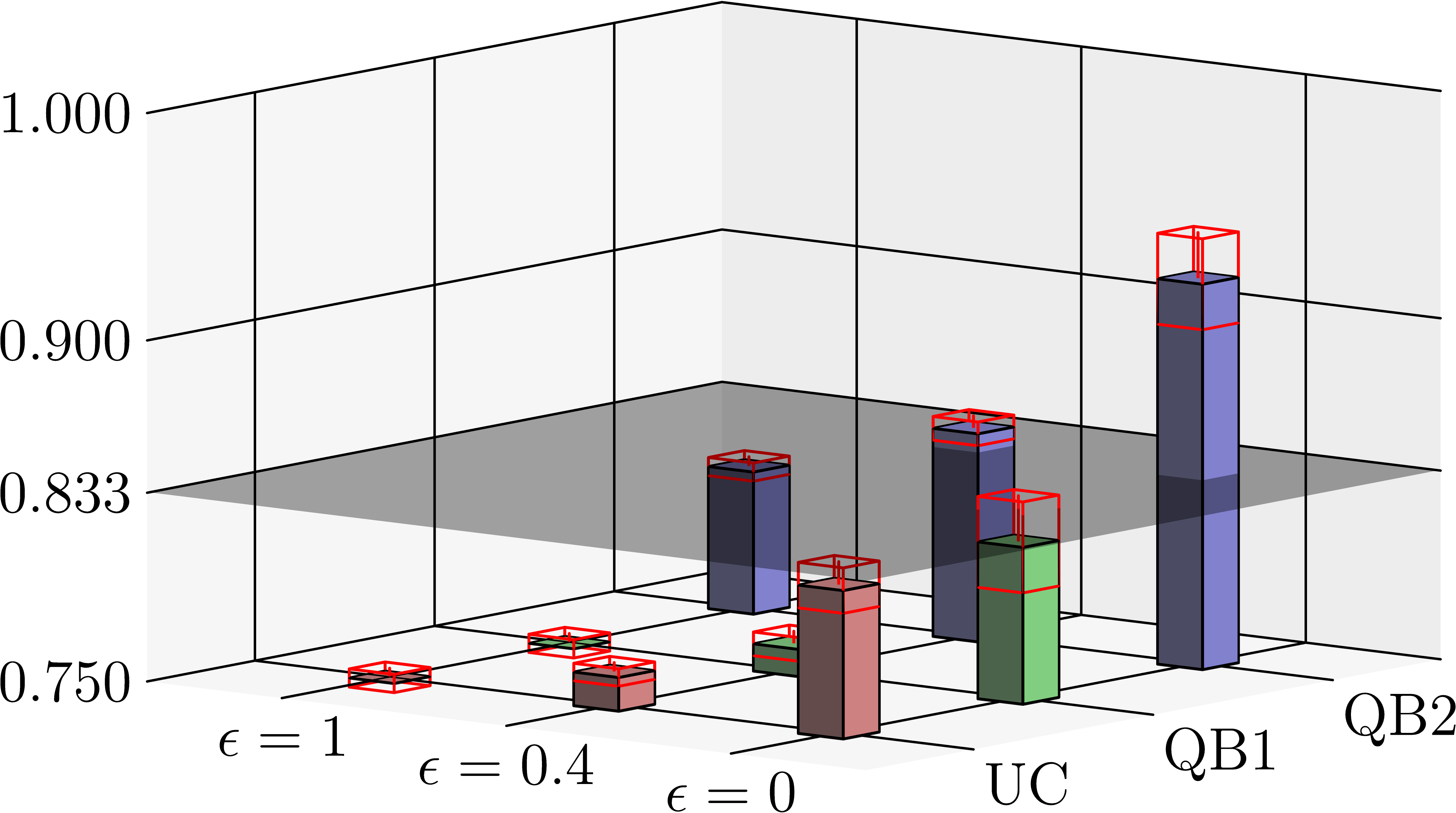}
\caption{\label{fig:results} Experimentally measured average cloning
fidelity $\bar{F}_i$ for quantum banknotes~1 and 2 (QB1 and QB2), 
and optimal universal cloning (UC). The red frames show the error bars of the measured probabilities. The verification threshold (gray surface) is set at $0.833$,  which is the fidelity of the optimal universal cloning process, \ie, the process  that ignores any information about $g$. }
\end{figure}

\noindent {\bf Experimental quantum forgery.} Let us consider
cloning the quantum banknote~1 from Fig.~\ref{fig:token}, where
single-photon polarization states appear approximately with the
following probabilities: $p(\D)=   0.125,$ $p(\A)=   0.125,$
$p(\V)=   0.125,$ $p(\Hl)=  0.125,$ $p(\Lc)=  0.25,$ $p(\R) =
0.25,$ where the poles of the Bloch sphere correspond to the
left-circular ($\Lc$)  and right-circular ($\R$) polarization
states, while the equatorial plane is spanned by the horizontal
($\Hl$), vertical ($\V$), diagonal ($\D$), and anti-diagonal
($\A$) polarization states. In this case the optimal cloning
machine is an axially-symmetric phase-covariant
cloner\cite{Karol10} corresponding to the
MPCC\cite{Karol09}. The probability distribution is described
here with only one nonzero number, \ie, $c_{2,0}=0.25\sqrt{5\pi}$
(using the notation from Ref.~{[\!\!\citenum{Karol10}]}:
$a_2=c_{2,0}/\sqrt{5\pi}$ and $|\Gamma|=0$). The fidelity of
copying the equatorial states is then equal to $F(\Hl)= F(\V) =
F(\D) = F(\A)=0.789$ and $F(\Lc) = F(\R) = 0.894$ for the pole
states. This results in the theoretical value of
$\bar{F_i}(\kappa\to\infty)=0.842$, which is a bit above the
security threshold of $\bar{F}_i=0.833.$ Using our experimental
setup shown in Fig.~\ref{fig:setup}, we achieve
$\bar{F}_{i,\mathrm{experiment}}= (81.9 \pm 2.0)\%$. This
experimental value is close to the universal cloning limit, \ie,
$\bar{F}_i=0.833$. In this case, only $(14.0 \pm 2.9)\%$ of the
sequence was  successfully copied. Alternatively, when we attack
this banknote with our implementation of the optimal universal
cloner, we obtain $\bar{F}_{i,\mathrm{experiment}}= (81.5 \pm
1.2)\%$; and $(19.6 \pm 1.2)\%$ of qubits are copied. This makes
the forgery unsuccessful for two reasons: (i) the quality of the
delivered qubits is lower than allowed, (ii) we delivered less
than $50\%$ of the sequence to each recipient. More than $50\%$ of
the qubits have to be delivered to exclude the possibility of
duplicating the money by cutting it into pieces. However, the
forgery becomes successful if one uses the optimal quantum cloning
process, with high fidelity but low success rate, 
interchangeably with a classical cloning process, with high success
rate but low fidelity.

Let us consider another case, where we can crack the  QMS and
the quantum banknote 2 from Fig.~\ref{fig:token} is described with
the following probabilities: $p(\D)=0.125,$ $p(\A)= 0.125,$
$p(\V)=0.125,$ $p(\Hl)=0.125,$ $p(\circlearrowleft) = 0.50,$ and
$p(\circlearrowright)= 0.$ In this case the optimal cloning
machine is also an axially-symmetric (phase-covariant)
cloner\cite{Karol10} (ASC), where $c_{1,0}=0.5\sqrt{3\pi},$
$c_{2,0}=0.25\sqrt{5\pi},$ which corresponds to  $a_1=0.5,$
$a_2=0.25,$ and $|\Gamma|=\infty,$ using the notation from
Ref.~{[\!\!\citenum{Karol10}]}. We have falsified this banknote by
applying interchangeably both the optimal classical  and the best
quantum copying strategies (see
Ref.~{[\!\!\citenum{Karol14amp}]}). The optimal classical copying
can be viewed as measuring a fraction $\epsilon$ of the original
photons from the sequence in a random basis (selected according to $g$)
and preparing two photons in the detected state. We implemented this
strategy by randomly swapping a fraction of photons from the
original sequence with the circularly-polarized photons selected
in accord with $g$ (for details see the Methods). The fidelity of
this strategy is $(3+\langle\cos\theta \rangle^2)/4=(3+a_1^2)/4$.
We used this optimal classical strategy with probability
$\epsilon = 0.4$. Using this method, we implemented a cloning
attack, which copies circa $(54.9\pm 0.1)\%$ of the sequence (this
means that we could sacrifice about $4\%$ of the sequence to
estimate $g$). Our implementation of the optimal quantum copying
strategy allows us to copy $24.8 \pm 0.1\%$ of the sequence with
a fidelity of $(92.4 \pm 0.4)\%$ (the theoretical value is
$92.6\%$). The  optimal classical copying
strategy\cite{Karol14amp} operated with fidelity circa $81.3\%$.
This provides us with the experimental average cloning fidelity of
$\bar{F}_{i,\mathrm{experiment}}=0.842\pm0.002$.
% about 0.36 of the quantum cloning RMS
Thus, we demonstrated that it is possible to crack the Wiesner QMS
with currently available technology. However, this was
possible only because the incoming sequence of photons was
synchronized with the probing photons allowing them to interact on
a beam splitter. The counterfeiter would face some additional
technical challenges when applying the discussed copying method in
real life (see the discussion in
Ref.~{[\!\!\citenum{Karol13prl}]}). This cloning regime, where the
cloning process  happens with a fidelity larger than the
fidelity of the best classical copying process, and the
transmitted qubits are successfully copied with a probability
larger than $50\%$, can also be  applied constructively to
increase the classical product capacity
of a quantum channel\cite{Karol14amp}.

The experimental results of the above-discussed copying
strategies for the two experimental quantum banknotes are
summarized in Fig.~\ref{fig:token2}.
Moreover, in Figs.~\ref{fig:probs}-\ref{fig:results}
we demonstrate how the  measured success probability of the cloning 
process and the corresponding single-copy fidelity depend on the value of 
the  hybridization parameter $\epsilon$. The selected values of this parameter 
correspond to optimal classical ($\epsilon=1$), hybrid ($\epsilon=0.4$), 
and  optimal quantum cloning ($\epsilon=0$). 
The significant reduction of
variance in these figures with respect to purely quantum cloning ($\epsilon=0$) 
is caused by using a robust classical copying process interchangeably with 
a more delicate optimal quantum cloning strategy (for details see the Methods). 
\\

%\section{Conclusions}
\noindent {\bf Conclusion.} We demonstrated that using
currently available technology we are able to both implement and
crack the original QMS of Wiesner\cite{Wiesner83}, given that
(i) a sequence of qubits, representing the quantum banknote is not
sampled uniformly over the Bloch sphere, (ii) the banknote is
considered valid if more than $50\%$ of the sequence is delivered
and its average fidelity is above the fidelity of  the
universal cloner\cite{UC96}, \ie, $83.3\%$. 
From our results it follows that to make 
the Wiesner QMS secure against copying, one should apply
a $g$-dependent verification threshold,
which corresponds to the average single-copy fidelity of the relevant 
optimal quantum cloner.
We  have shown that a
specialized  optimal cloner for an arbitrary qubit distribution $g$
can easily be  found by computing only its five parameters and
subsequently applying the optimization procedure described in
\mbox{Ref.~{\citeref{F1}}}. We believe that our results will
stimulate further research on secure quantum
communication and quantum technologies.\\

%\section*{Methods}
\noindent {\large\bf \\ Methods.}\\

%\subsection{Theory}
\noindent {\bf Theory.} In our theoretical considerations we apply
the spherical harmonics\cite{Arfken85}
$Y^m_l$ for $l=0,1,2$ and $m=0,1,...,l$. 
The spherical harmonics for $m<0$ are simply
related to these for $m>0$, because
%\begin{eqnarray}
%Y_{0,0} & = & \frac{1}{2\sqrt{\pi}}\\
%Y_{1,0} & = & \frac{1}{2}\sqrt{\frac{3}{\pi}}\text{cos}\theta\\
%Y_{1,1} & = & -\frac{1}{2}e^{i\phi}\sqrt{\frac{3}{2\pi}}\text{sin}\theta\\
%Y_{2,0} & = & \frac{1}{4}\sqrt{\frac{5}{\pi}}\left(-1+3\text{cos}^{2}\theta\right)\\
%Y_{2,1} & = &- \frac{1}{2}e^{i\phi}\sqrt{\frac{15}{2\pi}}\text{cos}\theta\sin\theta\\
%Y_{2,2} & = &- \frac{1}{4}e^{2i\phi}\sqrt{\frac{15}{2\pi}}\text{sin}^{2}\theta
%\end{eqnarray}
\begin{equation}\label{eq:prop}
Y^m_l=\left(-1\right)^{m}\bar{Y}^{-m}_l.
\end{equation}
The operator $\hat{R}$, in terms of the spherical harmonics
$Y_{l,m}$, can be expressed as
\begin{equation}
\hat{R}=\sum_{l=0}^{2}\sum_{m=-l}^{l}\hat{K}_{l,m}c_{l,m}\,,
\label{R}
\end{equation}
where
\begin{equation}
\hat{K}_{l,m}=\frac{1}{2}\int_{\Omega}\hat{\rho}^{\mathrm{T}}\otimes\left(\hat{\openone}\otimes\hat{\rho}+\hat{\rho}\otimes\hat{\openone}\right)\bar{Y}^m_l(\theta,\phi)\,\mathrm{d}\Omega\,,
\label{K_lm}
\end{equation}
the bar denotes complex conjugation, and
\begin{equation}
c_{l,m}=\int_{\Omega}g(\theta,\phi)Y^m_l(\theta,\phi)\,\mathrm{d}\Omega.
\end{equation}
It can  be  directly shown that
\begin{equation}
\hat{\rho}^{\mathrm{T}}\otimes\left(\hat{\openone}\otimes\hat{\rho}+\hat{\rho}\otimes\hat{\openone}\right)=2\sum_{l=0}^{2}\sum_{m=-l}^{l}\hat{K}_{l,m}Y^m_l(\theta,\phi)\,
,
\end{equation}
hence, we do not need terms with $l>2$. For a real-valued
distribution $g$ we obtain
\begin{equation}
c_{l,m}=\left(-1\right)^{m}\bar{c}_{l,-m}.
\end{equation}
This property follows from the definition of the spherical
harmonics. Thus, for the normalized $g$ distributions one computes
$c_{l,m}$ only for $l=1,2$ and $m=0,1,...,l$, which results in
five integrals in total. Depending on the symmetry of the
distribution $g$, some of the integrals vanish, which simplifies
further calculations. The expansion coefficients
$\hat{K}_{l,m}$ can be written in the form of block matrices
as given in the Supplementary Material\cite{Supplement}.

%\subsection{Experiment}
\noindent {\bf Experiment.} The experimental setup is depicted in
Fig.~\ref{fig:setup}. Pairs of photons were generated in the
process of spontaneous parametric down-conversion using a LiIO$_3$
crystal pumped by 200\,mW of cw Kr$^+$ laser beam at 413\,nm.
Hundreds of photon pairs were collected using single-mode
fibres and transferred to the input of the cloner setup. One
photon of each pair (\ie, a cloned photon) was used to encode a
bit of quantum information into its polarization state, while the
other photon served as an ancilla being either horizontally or
vertically polarized. In the next step, the cloned and ancillary
photon interfere on a polarization-dependent beam splitter (PDBS).
Ideally, this beam splitter should transmit the
horizontally-polarized light with intensity transmissivity of
0.789 and the vertically-polarized light with intensity
transmissivity of 0.211. Due to manufacturing errors, the real
intensity transmissivities of our PDBS are 0.76 and 0.18 for
horizontal and vertical polarisations, respectively. To correct
for this deviation between the real and ideal PDBS parameters, a
beam divider assembly (BDA) is inserted into each output mode of
the PDBS. This BDA consists of two beam displacers separating and
subsequently rejoining horizontal and vertical polarization
components of photons wave packets. By inserting a neutral-density
filter (NDF) into either a horizontal or vertical polarization mode
inside the BDA, one can achieve polarization sensitive losses and,
thus, compensate for incorrect parameters of the PDBS. Note that
this compensation can restore an ideal operation of the PDBS at
the expense of a lower success rate. To balance the rate of  the
cloned and ancillary photons, some additional NDFs can
be placed behind the BDAs. Finally, both the cloned and ancillary
photons are subjected to our polarization analysis consisting of a
set of quarter-wave  (QWP) and half-wave (HWP) plates followed by
a polarizing prism\cite{halenkova2012tom}. The coincident photon
detections are counted for each combination of the polarization
projection onto the horizontal, vertical, diagonal, anti-diagonal,
and both circular polarisations. The density matrices of the
corresponding two-photon states are then estimated using a
maximum-likelihood algorithm\cite{Jezek03}. 
A more detailed account on the
experimental procedure is available in our technical
paper\cite{Karel12}. The swapping procedure used for the optimal
classical copying strategy was implemented with the setup shown in
Fig.~\ref{fig:setup} by removing the PDBS and filters used in the
BDAs. We applied the following hybrid quantum-classical
cloning procedure: Initially, we prepared the best classical replacement
for $\rho=|\psi\rangle\langle\psi |$, \ie,
$\hat \sigma = \int_\Omega g |\psi\rangle\langle\psi | \, d \Omega$
in the ancillary mode and randomly swapped it with the input state $\hat{\rho}$ for
a fraction $\epsilon$ of the input photons. For the remaining $1-\epsilon$
photons we performed the relevant optimal quantum cloning.
When properly tuned, this procedure is far less noisy than the
implementation of pure quantum cloning and, thus, the
quality (described by, e.g., the dispersion of the fidelity) of
this hybrid cloning procedure depends mostly on the quality of
the quantum cloning process (see Figs.~\ref{fig:probs}-\ref{fig:results}).
%\bibliographystyle{naturemag}
%\bibliography{article}

%------------------------------------------------------------------
%\end{thebibliography}

\section*{\\Addendum}

\textbf{Acknowledgments.} We gratefully acknowledge the financial
support of the Polish National Science Centre under grant
DEC-2013/11/D/ST2/02638 (K.B., K.L.) and the support by the
GA\v{C}R grants No. 16-10042Y (K.B., K.L.) and No. P205/12/0382
(A\v{C}). F.N. is
partially supported by the RIKEN iTHES Project, MURI Center for
Dynamic Magneto-Optics, JSPS-RFBR contract no. 12-02-92100,
JST-IMPACT, and a Grant-in-Aid for Scientific Research (A).

\textbf{Correspondence.} Theoretical aspects corresponding author
K.B.~(email: bark@amu.edu.pl). Experimental aspects corresponding
authors A.\v{C}~(email: acernoch@fzu.cz) and K.L~(email:
k.lemr@upol.cz).

\newpage
\newpage
\begin{center}
{\Large Supplementary Material}
\end{center}

{\bf Abstract:} In this supplementary material we provide
more details on our theoretical approach. We recall definitions
and some properties of a few known optimal axially-symmetric
quantum cloners, which we have implemented experimentally in this
work. These include the universal cloner (UC), the phase-covariant cloner (PCC), and the
mirror-phase-covariant cloner (MPCC). We also present the
expansion coefficients $\hat{K}_{l,m}$. 
Moreover, we present an additional figure of the measured fidelities.

\section*{Axially-symmetric quantum cloning}

\begin{figure}
\includegraphics[width=8.5cm]{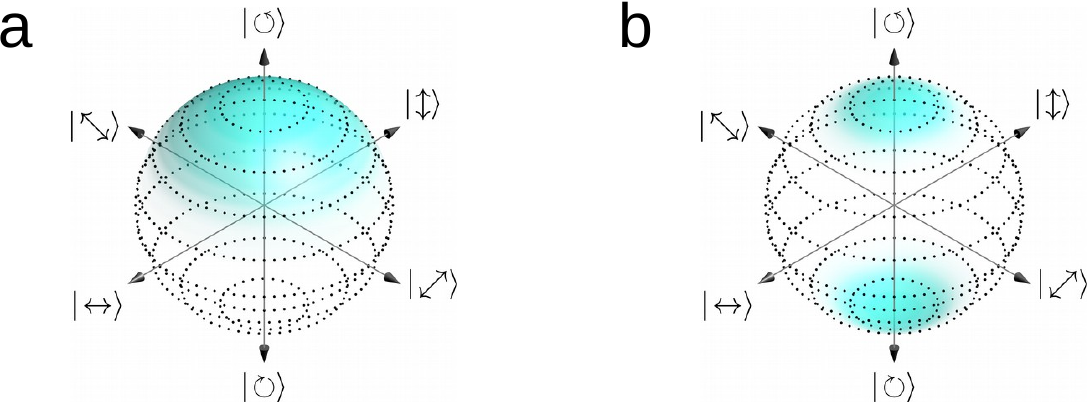}
\caption{\label{fig:bloch} Assorted axially-symmetric qubit
distributions over the Bloch sphere, where the shade corresponds
to the probability of choosing a qubit at the specific point on
the sphere. The first example (a) is an axially-symmetric qubit
distribution, which is optimally cloned by PCC. The next example
(b), corresponds to an axially-symmetric qubit distribution with
additional mirror symmetry. This distribution is optimally cloned
by MPCC.}
\end{figure}

The figure of merit for the quantum cloning machines is  the
fidelity of  their clones. The special case, where the qubits
are uniformly distributed around the poles of the Bloch sphere
corresponds to the axially-symmetric cloning described in
Ref.\citeref{Karol10}. This class of cloning machines includes
both phase-covariant cloners (PCC) and mirror-phase-covariant
cloners (MPCC) as special cases. The former
is a cloning process optimized for a given qubit
distribution, where there is a higher chance of cloning a qubit
corresponding to one of the poles than the other pole (see
Fig.~\ref{fig:bloch}). The latter optimal cloning process is
optimized for mirror-symmetric distribution on a Bloch sphere (see
Fig.~\ref{fig:bloch}). Note that the optimal universal cloner (UC)
is a special case of MPCC.

An arbitrary optimal $1\to2$ cloning of qubits given by an
axially-symmetric distribution can be expressed as a unitary
transformation\cite{Karol10}
\begin{eqnarray}
\ket{\Lc}_{a}\ket{\Lc\R}_c&\to& \Lambda_+\ket{\Lc\Lc}_{a,b}\ket{\R}_c + \bar{\Lambda}_+\ket{\,\psi}_{a,b}\ket{\Lc}_c,\quad\\
\ket{\R}_{a}\ket{\Lc\R}_c&\to&
\Lambda_-\ket{\R\R}_{a,b}\ket{\Lc}_c +
\bar{\Lambda}_-\ket{\,\psi}_{a,b}\ket{\R}_c,\quad
\end{eqnarray}
where $\bar\Lambda_\pm = \sqrt{1-\Lambda_\pm^2}$,
$\ket{\,\psi}=(\ket{\Lc\R}+\ket{\R\Lc})/\sqrt{2}$, and $a,b,c$
stand for the two copy modes and the ancillary mode, respectively.
For the PCC, $\Lambda_\pm \in \{0,1\}$ and
$\Lambda_+=1-\Lambda_-$. For example, if $p(\Lc)\gg p(\R)$, then
$\Lambda_+=1$ and $\Lambda_-=0$. In particular, the cloning
transformation for the mirror-phase-covariant cloner\cite{Karol09}
can be expressed as
\begin{eqnarray}
\ket{\Lc}_{a}\ket{\Lc\R}_c&\to& \Lambda\ket{\Lc\Lc}_{a,b}\ket{\R}_c + \bar{\Lambda}\ket{\,\psi}_{a,b}\ket{\Lc}_c,\\
\ket{\R}_{a}\ket{\Lc\R}_c&\to& \Lambda\ket{\R\R}_{a,b}\ket{\Lc}_c
+ \bar{\Lambda}\ket{\,\psi}_{a,b}\ket{\R}_c.
\end{eqnarray}
Quantum banknote~1 is cloned in the optimal way if $\Lambda =
0.88$. The optimal universal cloning (UC) is a special case of
the MPCC and it corresponds to $\Lambda =\sqrt{2/3}$.

In  the case of an axially-symmetric distribution, where all
equatorial states appear with the same probability, there is a
single parameter\cite{Karol10}
\begin{equation}
\Gamma =
\frac{\sqrt{2}\gamma_-(\gamma_+-1)}{\gamma_+^2-\gamma_-^2},
\end{equation}
that can be used to select the optimal cloning transformation,
which in the case of quantum banknotes~1 and 2 depends only on the
probabilities $p(\Lc)$ and $p(\R)$, because $\gamma_\pm =
p(\Lc)\pm p(\R)$. If $|\Gamma|>1$, the optimal cloner is the 
PCC. Alternatively, if $|\Gamma|=0$, the optimal cloner is the
MPCC. Here, we analyse the case where $p(\Lc)= p(\R)$ or
$p(\R)=0$.

\section*{Hybrid  quantum-classical cloning}

\begin{figure}

\includegraphics[width=8.5cm]{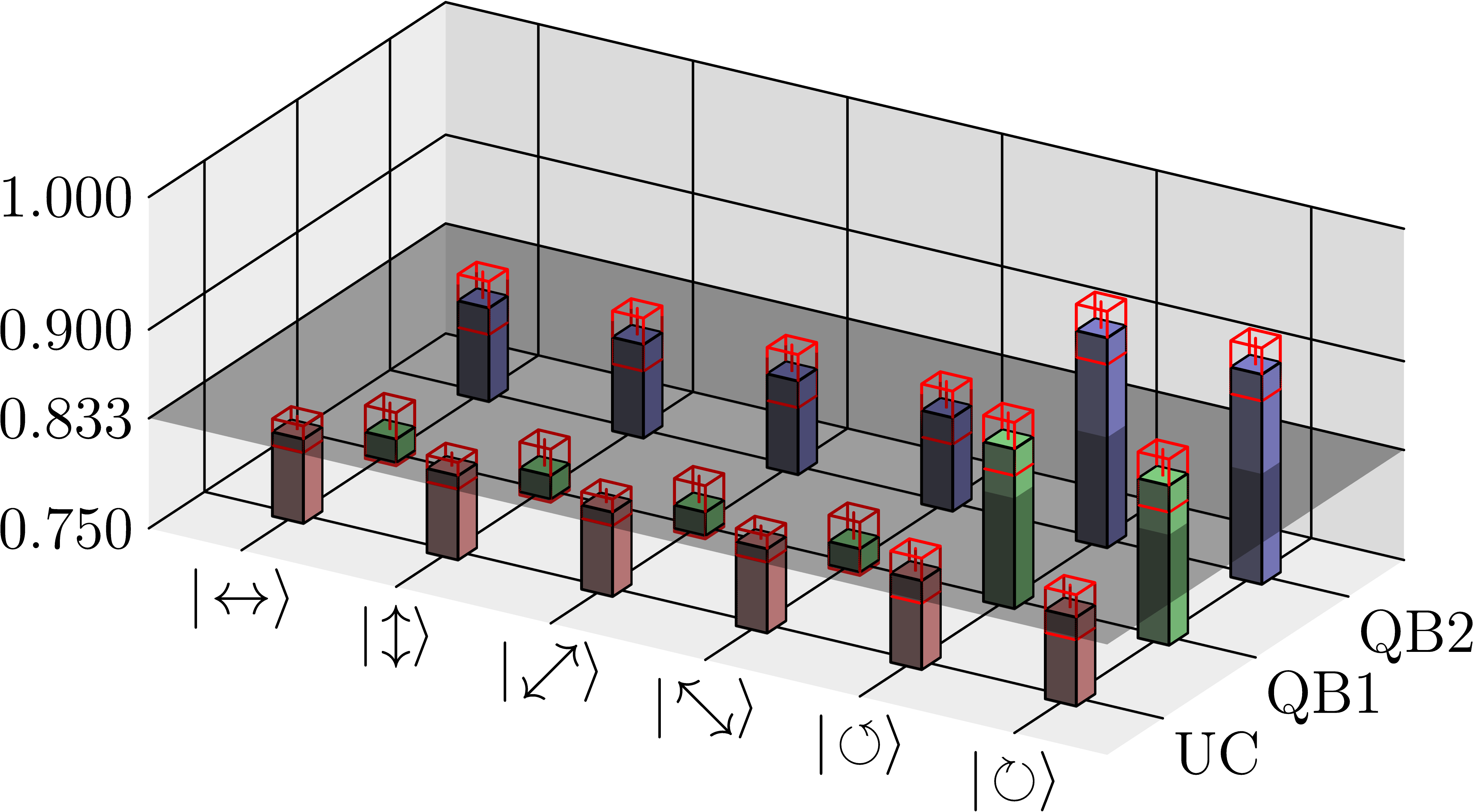}
\caption{\label{fig:fidelities} Single-copy
fidelities $\bar F_i$ for the states with different linear and
circular polarizations measured during the encoding of our two
quantum banknotes 1 and 2 (QB1 for $\epsilon=0$ and QB2 for
$\epsilon=0.4$), and the optimal universal cloner (UC  for $\epsilon=0$). 
The grey surface shows the theoretical fidelity of the universal cloner. The 
red frames show the error bars of the measured fidelities. 
This figure can be compared with Figs. 5 and 6 in the main article.}
\end{figure}

Our linear-optical implementation of the quantum cloning
machine works probabilistically. Thus, to improve the cloning
efficiency, we apply, interchangeably, the optimal quantum and
deterministic classical copying processes. Specifically, we classically (or quantumly) copied a fraction $\epsilon$
(or $1 - \epsilon$) of the input photons. Thus, $\epsilon$ can be
treated as a hybridization parameter.
The measured fidelities are presented in Fig~\ref{fig:fidelities}. 

\begin{figure*}
\includegraphics[width=14.5cm]{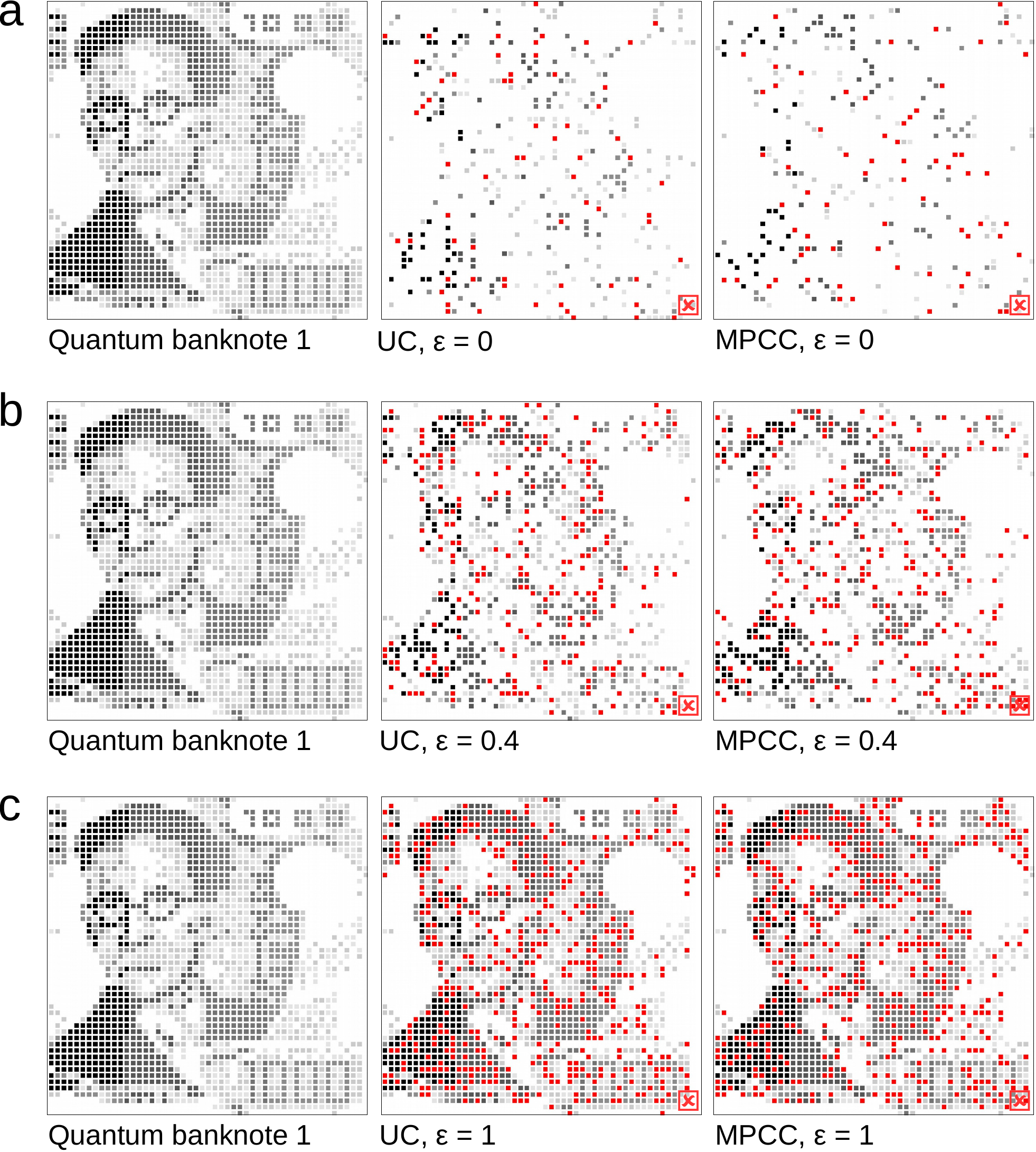}
\caption{ \label{fig:token3} Experimental quantum banknote
1 is copied probabilistically, and subsequently verified, 
 with the optimal $1$-to-$2$ 
linear optical cloning machine shown in
Fig.~\ref{fig:setup} . 
One observes that the copies, which are
provided with the best possible cloning machines, are noisy and,
thus, the sequences of qubits are damaged (shown in red). The
performance of the cloning process depends on the statistics of
photon polarisations and on the hybridisation  parameter $\epsilon$,
as shown in panels (a), (b), and (c).}
\end{figure*}

\begin{figure*}
\includegraphics[width=14.5cm]{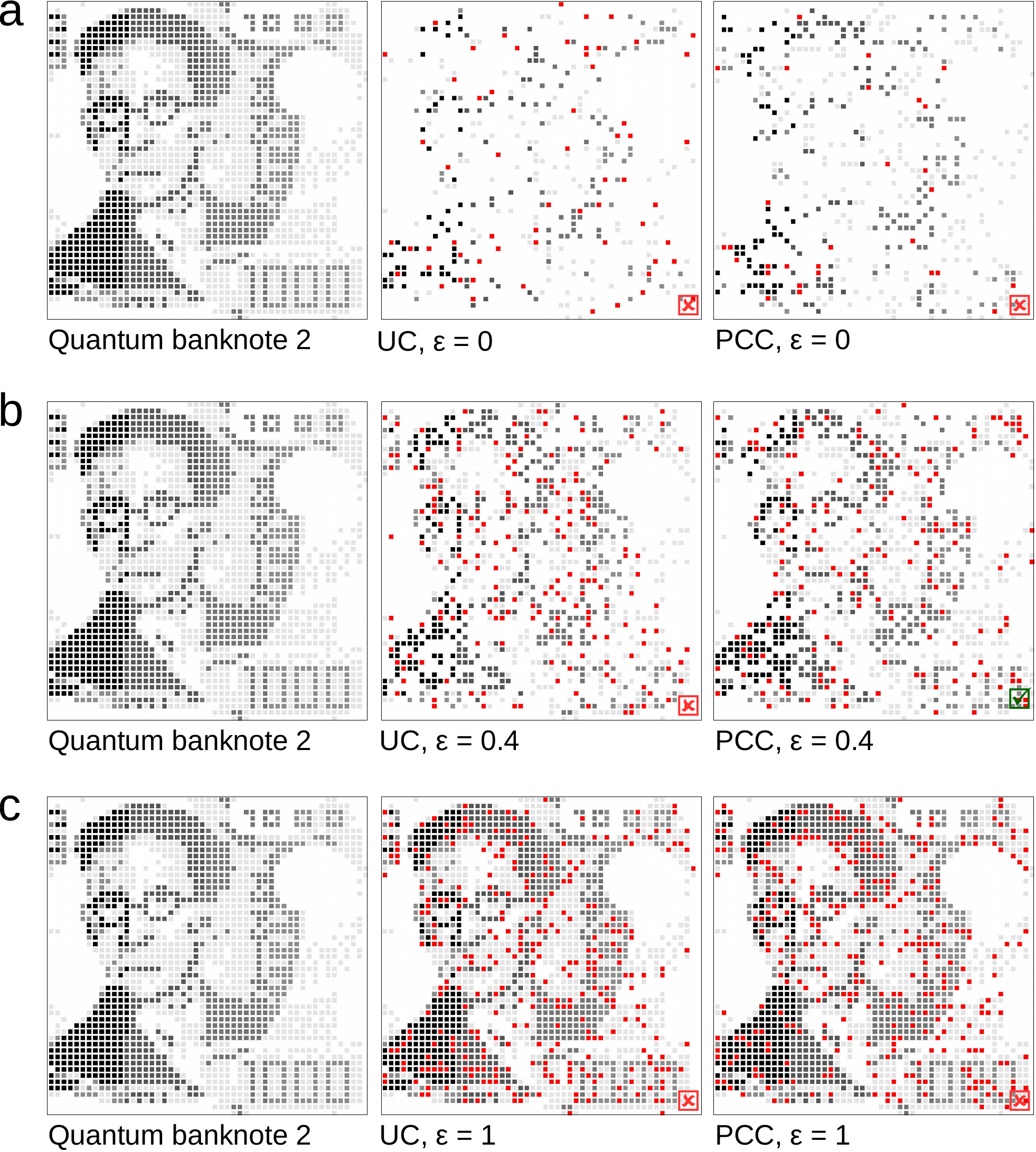}
\caption{ \label{fig:token4} Same as in Fig.~\ref{fig:token3} but
for quantum banknote 2.}
\end{figure*}
\vspace*{1cm}
\section*{Expansion coefficients $\hat{K}_{l,m}$}

The expansion coefficients $\hat{K}_{l,m}$, given in
Eq.~(\ref{K_lm}), of the operator $\hat R$, given in
Eq.~(\ref{R}), can be written in the form of block matrices as
follows:
\begin{equation}
\hat{K}_{0,0}=\frac{\sqrt{\pi}}{12}\left(\begin{array}{c|c} 3\hat A+
\hat B & 2\hat C^\mathrm{T}\\\hline 2\hat C & 3\hat A-\hat
B
\end{array}\right)\,,
\end{equation}

\begin{equation}
\hat{K}_{1,0}=\frac{1}{2}\sqrt{\frac{\pi}{12}}\left(\begin{array}{c|c}
\hat{A}+\hat{B} & \hat 0\\\hline \hat 0 & -\hat A+ \hat B
\end{array}\right)\,,
\end{equation}

\begin{equation}
\hat{K}_{1,1}=-\sqrt{\frac{\pi}{24}}\left(\begin{array}{c|c} \hat
C & \hat A\\\hline \hat 0 & \hat C
\end{array}\right)\,,
\end{equation}

\begin{equation}
\hat{K}_{2,0}=-\frac{1}{6}\sqrt{\frac{\pi}{5}}\left(\begin{array}{c|c}
-\hat B & \hat C^\mathrm{T}\\\hline \hat C & \hat B
\end{array}\right)\,,
\end{equation}

\begin{equation}
\hat{K}_{2,1}=-\frac{1}{2}\sqrt{\frac{\pi}{30}}
\left(\begin{array}{c|c} \hat C & \hat B\\\hline \hat 0 & -\hat C
\end{array}\right)\,,
\end{equation}

\begin{equation}
\hat{K}_{2,2}=\sqrt{\frac{\pi}{30}} \left(\begin{array}{c|c} \hat
0 & \hat C\\\hline \hat 0 & \hat 0
\end{array}\right)\,,
\end{equation}
where $\hat A = \mbox{diag}[2,2,2,2]$, $\hat B =
\mbox{diag}[2,0,0,-2]$, $C_{ij} =
(\delta_{i,2}+\delta_{i,3})\delta_{j,1} +
(\delta_{j,2}+\delta_{j,3})\delta_{i,4}$, and $\hat 0$ is a $4\times 4$ matrix of
zeros. Moreover, it follows from Eq.~(\ref{eq:prop}) that
$\hat{K}_{l,m}=\left(-1\right)^{m}\hat{K}_{l,-m}^{\mathrm{T}}.$

\end{document}